\documentclass[aps, pra, reprint, showpacs, groupedaddress, superscriptaddress, amsfonts, amsmath, amssymb]{revtex4-1}

\usepackage{dcolumn} 
\usepackage{bm} 
\usepackage{graphicx}
\usepackage{braket}
\usepackage{hyperref}
\usepackage{units}
\usepackage{dsfont} 
\usepackage{xcolor}

\definecolor{darkblue}{rgb}{0, 0, 0.8}
\hypersetup{
	pdftitle={Topologically protected edge states in small Rydberg systems}, 
	pdfauthor={Sebastian Weber},      
	pdfsubject={Quantum Physics},  
	pdfcreator={Sebastian Weber},     
	pdfproducer={Sebastian Weber},    
	colorlinks=true,
	linkcolor=darkblue,
	citecolor=darkblue,
	filecolor=darkblue,
	urlcolor=darkblue
}

\begin{document}

\title{Topologically protected edge states in small Rydberg systems}

\author{Sebastian Weber}
\email{weber@itp3.uni-stuttgart.de}
\affiliation{Institute for Theoretical Physics III and Center for Integrated Quantum Science and Technology, University of Stuttgart, 70550 Stuttgart, Germany}
\author{Sylvain de L\'es\'eleuc}
\author{Vincent Lienhard}
\author{Daniel~Barredo}
\author{Thierry Lahaye}
\author{Antoine Browaeys}
\affiliation{Laboratoire Charles Fabry, Institut d'Optique Graduate School, CNRS, Universit\'e Paris-Saclay, 91127 Palaiseau Cedex, France}
\author{Hans Peter B\"uchler}
\affiliation{Institute for Theoretical Physics III and Center for Integrated Quantum Science and Technology, University of Stuttgart, 70550 Stuttgart, Germany}

\begin{abstract}
We propose a simple setup of Rydberg atoms in a honeycomb lattice which gives rise to topologically protected edge states. The proposal is based on the combination of dipolar exchange interaction, which couples the internal angular momentum and the orbital degree of freedom of a Rydberg excitation, and a static magnetic field breaking time reversal symmetry. We demonstrate that for realistic experimental parameters, signatures of topologically protected edge states are present in small systems with as few as 10 atoms. Our analysis paves the way for the experimental realization of Rydberg systems characterized by a topological invariant, providing a promising setup for future application in quantum information.
\end{abstract} 

\maketitle

\textit{Introduction.} Systems characterized by topological invariants give rise to many interesting phenomena \cite{Hasan_2010, Qi_2011}. Of special significance are topologically protected edge states which arise in finite systems as the characteristic feature of topological band structures \cite{Halperin_1982, Essin_2011}. Topologically protected edge states possess distinguished properties like robustness to local perturbations which make them highly interesting for various applications such as the processing or coherent transport of quantum information \cite{Nayak_2008, Lang_2017, Dlaska_2017}. The prime example for the occurrence of topologically protected edge states is the integer quantum Hall effect \cite{Klitzing_1980, Laughlin_1981}. In addition, topological band structures and topologically protected edge states are observed in systems as varied as classical phononic \cite{Susstrunk_2015, Serra_Garcia_2018} or photonic \cite{Hafezi_2013, Lu_2016, Mili_evi__2017} setups, solid state systems \cite{Tsukazaki_2007, Konig_2007, Drozdov_2014}, and cold gases \cite{Jotzu_2014, Aidelsburger_2014, Stuhl_2015, Mancini_2015, Flaschner_2016}. For future application in quantum information or the realization of interesting many-body states on the basis of topological band structures \cite{Tsui_1982, Laughlin_1983,Tang_2011, Regnault_2011}, highly controllable quantum mechanical systems are required.

Due to recent experimental progress, Rydberg atoms have become a promising candidate for the realization of such systems.
The excellent experimental controllability of the long-ranging dipolar interaction between Rydberg atoms \cite{de2017accurate} and the capability of preparing arbitrary arrays of up to $\sim 50$ atoms \cite{Barredo_2016, Endres_2016, Kim_2016} paved the way for the realization of exotic matter \cite{Labuhn_2016, lienhard2017observing, Bernien_2017, kim2017detailed}. For polar molecules~\cite{Peter_2015} and recently, for photonic setups \cite{Bettles_2017, Perczel_2017}, it was suggested to realize topologically protected edge states by exploiting the intrinsic spin-orbit coupling of dipolar interactions in combination with broken time-reversal symmetry. We adapt this idea to mesoscopic systems of Rydberg atoms.

In this paper, we present a detailed proposal for a Rydberg system which gives rise to topologically protected edge states through dipolar interaction and provide experimentally feasible parameters for its implementation. Our proposal relies on an optical honeycomb lattice occupied by one Rydberg atom per lattice site. We assume that the atoms are in a Rydberg $\ket{nS}$ state except for one atom that is excited to a dipole-coupled $\ket{n'P}$ state. We show that in the presence of static electric and magnetic fields, which are isolating the relevant Rydberg state from the Rydberg manifold and where the later breaks time reversal symmetry, the band structure of such an excitation features non-zero Chern numbers. Note that in contrast to the original proposal for polar molecules \cite{Peter_2015} and setups relying on lattice shaking \cite{Hauke_2012,  Jotzu_2014, Flaschner_2016}, laser-assisted tunneling \cite{Miyake_2013, Aidelsburger_2013}, or synthetic dimensions \cite{Celi_2014, Stuhl_2015, Mancini_2015}, we do not need time-dependent fields for the realization of a topological interesting band structure. For studying edge states in finite systems, we develop a band structure analogue. Our analysis show that signatures of topologically protected edge states are present in systems with as few as $10$ atoms. Considering lattice disorder, we demonstrate in realistic systems that the edge states can be probed by observing the chiral movement of an edge excitation. The requirements for our proposal are readily met by recent Rydberg experiments.

\begin{figure}
\includegraphics{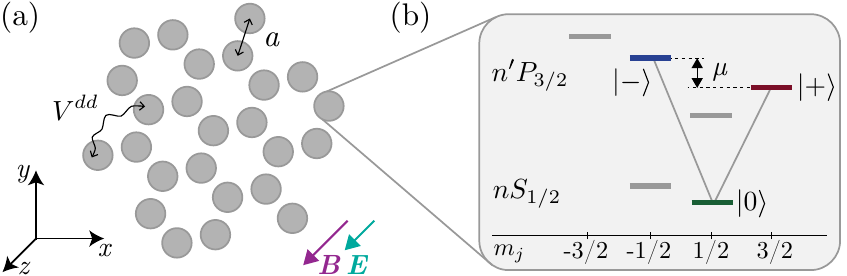}%
\caption{\label{fig:setup} Setup. (a) Each site of an optical honeycomb lattice is occupied by one Rydberg atom. The quantization axis $z$ is chosen perpendicular to the lattice. Static electric and magnetic fields are applied along $z$, to isolate the $V$-level structure (b). We consider all atoms to be in the state $\ket{0}$ except for one atom excited to the state $\ket{+}$ or $\ket{-}$. The excitation propagates through the system by means of dipole-dipole interaction.}
\end{figure}

\textit{Setup.} We consider an optical lattice occupied by one Rydberg atom per lattice site, see Fig. \ref{fig:setup}~(a). In this paper, we use a honeycomb lattice, but our proposal would work similarly for other lattice geometries such as square or Kagome lattices. We chose the quantization axis to be perpendicular to the lattice. We are interested in a V-level structure comprising the Rydberg states $\ket{0}= \ket{nS_{1/2}, m_j=1/2}$, $\ket{+}=\ket{n'P_{3/2}, m_j=3/2}$, and $\ket{-}=\ket{n'P_{3/2}, m_j=-1/2}$, see Fig. \ref{fig:setup}~(b). We apply static, homogeneous electric and magnetic fields along the quantization axis to lift the Zeeman degeneracy to isolate the V-level structure from other Rydberg states. The energy difference $\mu=E(\ket{+}) - E(\ket{-})$ can be adjusted by the fields.

We regard the state where all atoms are in the state $\ket{0}$ as the ground state. In the following, we study systems containing one excitation to the state $\ket{+}$ or $\ket{-}$. The excitation propagates through the system by means of dipole-dipole interaction. Note that for our analysis, we just consider dipolar exchange interaction and neglect the static dipole-dipole interaction of the finite dipole moments of the Rydberg atoms induced by the electric field. In addition, we ignore van der Waals interaction. These approximations are justified in Appendix \ref{sec:comparison}. By describing the creation of a $\ket{\pm}$ excitation at lattice site $i$ by the operator $b_{i,\pm}^\dagger = \ket{\pm}_i \bra{0}_i$ and introducing the spinor  $\boldsymbol{\psi}_i^\dagger = (b_{i,+}^\dagger, b_{i,-}^\dagger)$, we can write the operator for the dipolar interaction between the lattice site $i$ and $j$ as~\cite{Peter_2015} 
\begin{equation}
V_{ij}^\text{dd} = \frac{a^3}{|\boldsymbol{R}_{ij}|^3} \boldsymbol{\psi}_i^\dagger
\begin{pmatrix}
- t_+ & w  \text{e}^{-2\text{i}\phi_{ij}} \\
w  \text{e}^{2\text{i}\phi_{ij}} & - t_- \\
\end{pmatrix}
\boldsymbol{\psi}_j
+
\text{h.c.}
\;,
\label{eqn:interaction}
\end{equation}
where $a$ is the lattice constant, $\boldsymbol{R}_{ij} = \boldsymbol{R}_j - \boldsymbol{R}_i$ is the distance vector between the lattice sites, and $\phi_{ij}$ is the angle between the distance vector and the $x$-axis. The parameters $t_+$ and $t_-$ are the amplitudes of the hopping processes which conserve the internal angular momentum of the excitation. The amplitude $w$ belongs to the hopping process that flips a $\ket{-}$ excitation  into a $\ket{+}$ excitation, leading to a change in internal momentum by two which is compensated by a change in orbital momentum accounted by be the phase factor $\text{e}^{-2\text{i}\phi_{ij}}$.  Note that in our case of long-range interaction, the phase factor cannot be obtained through gauge fluxes.

The total Hamiltonian of the system reads
\begin{equation}
H = \frac{1}{2} \sum_{i \neq j} \hat{V}_{ij}^\text{dd} + \sum_i \boldsymbol{\psi}_i^\dagger 
\begin{pmatrix}
\mu/2 & 0 \\
0  & - \mu/2 \\
\end{pmatrix}
\boldsymbol{\psi}_i
\;,\label{eqn:hamiltonian}
\end{equation}
where the last sum includes the energy difference between $\ket{+}$ and $\ket{-}$ excitations.

\textit{Topological band structure.} For broken time reversal symmetry, i.e. $\mu \neq 0$ or $t_+ \neq t_-$, the topological properties of this Hamiltonian are characterized by  Chern numbers \cite{Peter_2015, Ryu_2010}. Note that $t_+ \neq t_-$ is intrinsically fulfilled by our setup because of the different Clebsch-Gordan coefficients for the creation of $\ket{+}$ and $\ket{-}$ excitations.

\begin{figure}
\includegraphics{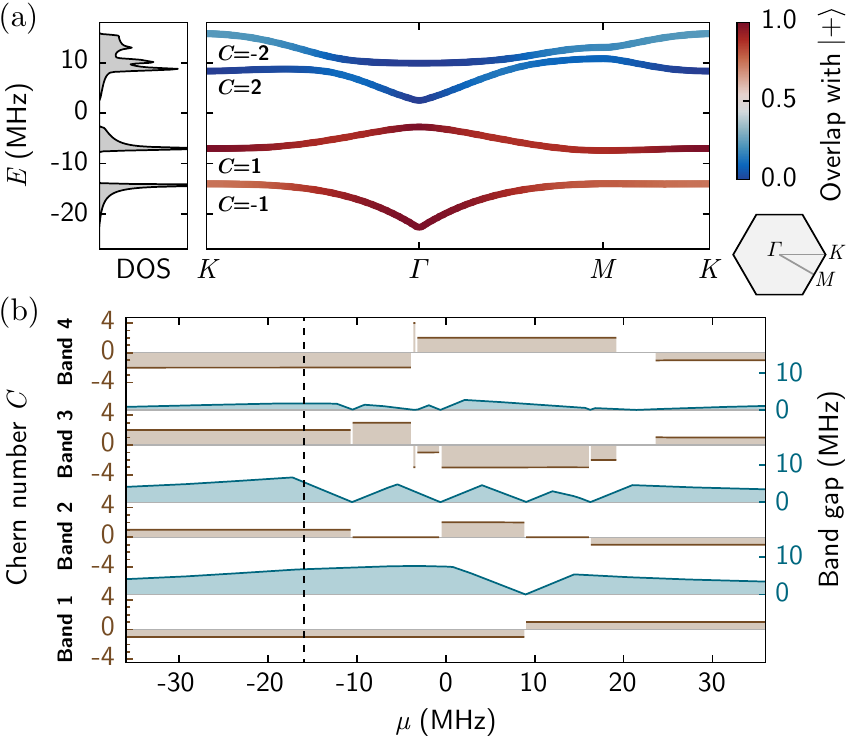}%
\caption{\label{fig:chernnumbers}(a) Density of states (DOS) and topological band structure of the infinite honeycomb lattice plotted over the depicted path through the Brillouin zone for a typical parameter set.  The bands are labeled with their Chern number $C$. The color code tells the overlap with the $\ket{+}$ state. (b) Chern numbers and band gaps as a function of $\mu$ for $w = \unit[4.17]{MHz}$, $t_+= \unit[2.41]{MHz}$, and $t_-= \unit[0.80]{MHz}$. The dashed line marks the parameter set used in (a) with $\mu = -\unit[16]{MHz}$ for which the Chern numbers of the two lower bands are highly robust against perturbations due to the pronounced band gaps. We use this parameter set for all calculations.}
\end{figure}
Fig. \ref{fig:chernnumbers} (a) shows the density of states and the topological band structure of the infinite honeycomb lattice for a typical set of parameters. As the unit cell of the honeycomb lattice consists of two sites and the $\ket{\pm}$ excitation has a two-fold internal degree of freedom, there are four bands in total. For each band, we calculate the Chern number $C$ \cite{bernevig2013topological, Fukui_2005} which depends on $\mu$ as shown in Fig.~\ref{fig:chernnumbers}~(b). Our system exhibits a rich phase diagram with Chern numbers ranging from $C=-4$ to $C=4$ as a function of $\mu$ for typical hopping amplitudes $w = \unit[4.10]{MHz}$, $t_+= \unit[2.25]{MHz}$, and $t_-= \unit[0.84]{MHz}$ whose experimental realization is discussed at the end of the paper. For further calculations, we use these hopping amplitudes together with $\mu=-\unit[16]{MHz}$. The selected parameters have the advantage that pronounced band gaps exist which make the Chern numbers $C=-1$ and $C=1$ of the two lower bands robust against perturbations. Moreover, because $\mu$ is much larger than $w$, the two lower bands mainly overlap with the $\ket{+}$ state what will turn out to be useful. In the following, we will focus on the two lower bands.

\begin{figure*}
\includegraphics{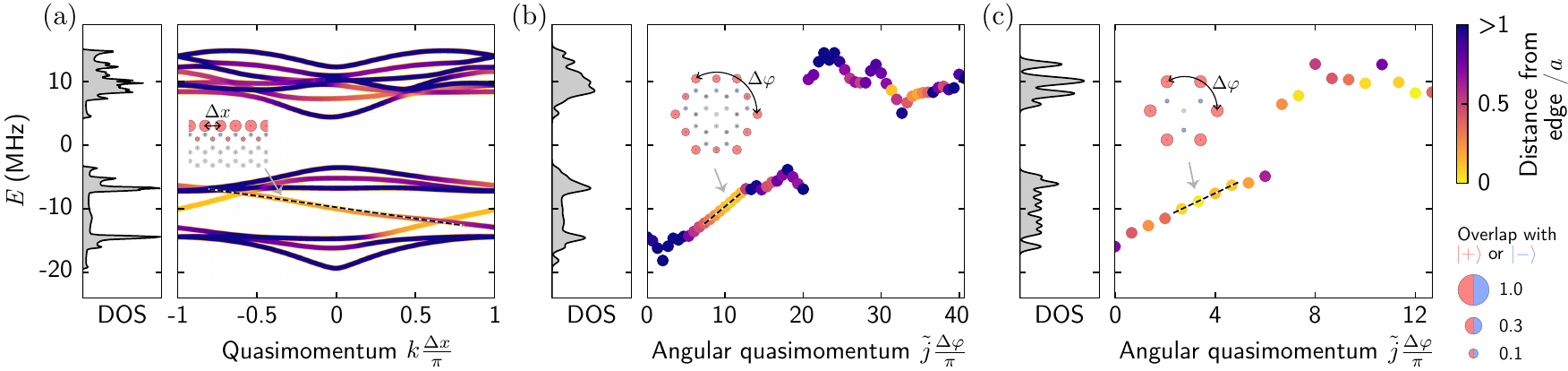}%
\caption{\label{fig:edgestates}(a) Density of states and topological band structure of a semi-infinite honeycomb lattice. The color code visualizes the expectation value of the distance of an excitation from the edge of the lattice, highlighting edge states which are the characteristic feature of the non-zero Chern numbers of the infinite system. The inset shows an exemplary edge state where the area of the red or blue dots is proportional to the probability of $\ket{+}$ or $\ket{-}$ excitations at a lattice site. The slope of the dashed line fitted to the edge mode is the group velocity $v_g = \unit[12.3]{a / \mu s}$ of an excitation at the upper edge. (b) Density of states convoluted by a narrow Gaussian and topological band structure of a disk-shaped system with 31 atoms plotted over the extended Brillouin zone. The angular group velocity of an edge excitation is $w_g = 2\pi \times \unit[0.68]{MHz}.$ (c) Signatures of topologically protected edge states are present even in a small disk-shaped system with as few as 10 atoms, where $w_g = 2\pi \times \unit[1.24]{MHz}$.}
\end{figure*}

\textit{Edge states.} In systems with boundaries, edge states are the characteristic feature of the topological bands of the infinite system as stated by the bulk-boundary correspondence \cite{Essin_2011}. We first study edge states in a semi-infinite system before heading towards experimentally feasible, small systems. Fig. \ref{fig:edgestates} (a) shows the band structure and density of state of an exemplary semi-infinite honeycomb lattice. For the eigenstates belonging to the bands, the color code visualizes the expectation value of the distance of an excitation from the edge.
The analysis shows that the gap, which was present between the two lower bands in the infinite system, is now closed by one chiral edge mode at the bearded edge of the considered semi-infinite system and one chiral edge mode at the zigzag edge. The dispersion relations of the edge modes are nearly linear. The inset of Fig. \ref{fig:edgestates} (a) shows an exemplary edge state.

As examples of experimentally well realizable systems, we study small disk-shaped systems with $10-31$ Rydberg atoms. The implementation of such systems is realistic considering recent experimental developments \cite{Barredo_2016, Endres_2016, Kim_2016}.
As in case of the semi-infinite system, we can calculate the density of states, see Fig. \ref{fig:edgestates} (b-c). The density of states shows that as before, edge states close the band gap which was existing in the infinite system, see Fig.~\ref{fig:edgestates}~(b-c). Yet, for analyzing edge states, the density of states is much less meaningful than the band structure. It neither tells the number of edge modes nor their dispersion. Therefore, we developed a band structure analogue which provides us this information. Contrary to similar approaches \cite{Hafezi_2013, Susstrunk_2015, Bettles_2017}, our band structure analogue allows us to make full use of the rotational symmetry of the considered disk-shaped systems. In the following, we explain the determination of the band structure analogue. Since the disk-shaped systems are $C_3$ symmetric, the eigenstates are Bloch waves which can be labeled by 
two quantum numbers: the band index $b$ and 
the \emph{total} angular quasimomentum $j\in\{0,1,2\}$
(the orbital angular quasimomentum is not a good quantum number because it is coupled to the internal angular momentum of the excitation).
\begin{figure}[b]
\includegraphics[width=\linewidth]{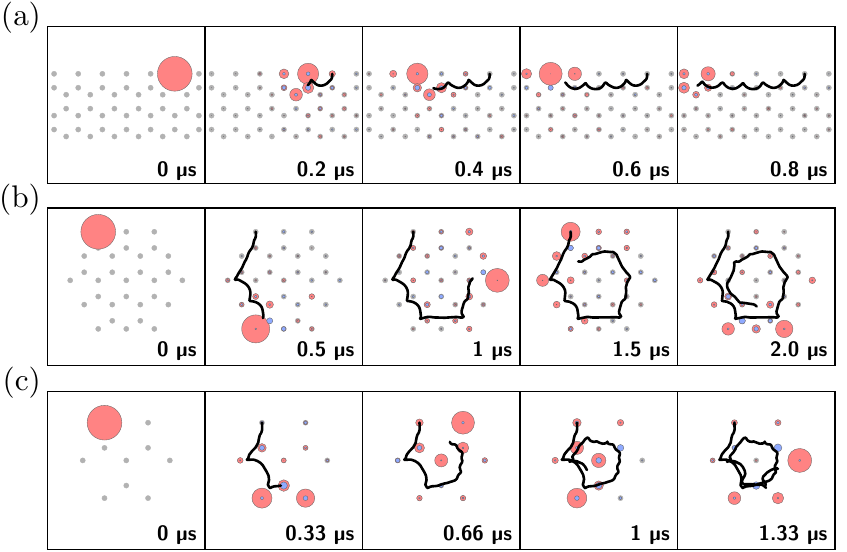}%
\caption{\label{fig:propagation} Chiral propagation of edge excitations into the $\ket{+}$ state, probing the edge states between the two lowest bands in the semi-infinite lattice (a), the disk-shaped system with $31$ atoms (b), and the disk-shaped system with $10$ atoms (c). The center of mass movements of the excitations are depicted as black lines. Their velocities agree with the group velocities extracted from the band structures shown in Fig. \ref{fig:edgestates}.}
\end{figure}
By Bloch's theorem, the coefficients of an eigenstate $\ket{\Psi_{b,j}}$ are
\begin{equation}
\braket{\pm|_i\; \Psi_{b,j}}
= \text{e}^{\text{i}j\varphi_i} \text{e}^{\mp \text{i}\varphi_i} \braket{\pm|_i\;  u_{b,j}}
\;,
\end{equation}
where $\varphi_i$ is the angular position of the $i$'th lattice site and $\ket{u_{b,j}}$ the $C_3$ periodic part of the Bloch wave.
The three possible values of $j$ make up the reduced Brillouin zone of the system. However, for analyzing edge states, we would like to plot the dispersion relation over a larger Brillouin zone because the edge states are exponentially localized at the edge of the disk-shaped systems which acquires a higher rotational symmetry than the bulk.
In order to show the $C_3$ symmetric bulk states as well as the higher symmetric edge states in one plot, we plot the band structure over the extended Brillouin zone. For this purpose, we have to uniquely assign to each of the $N$ eigenstates a quasimomentum $\tilde{j} \in \{0, ..., N-1\}$ from the extended Brillouin zone so that the energy of $\ket{\Psi_{\tilde j}}$ plotted over $\tilde{j}$ make up our band structure analogue.  As the periodic part $\ket{u_{\tilde j}}$ of a Bloch wave $\ket{\Psi_{\tilde j}}$ typically changes slowly with the quasimomentum $\tilde j$, the assignment should be such that
\begin{equation}
\sum_{\tilde{j}=0}^{N-2} | \braket{u_{\tilde j} |  u_{\tilde j+1} } |^2 = \sum_{\tilde{j}=0}^{N-2} | \braket{\Psi_{\tilde j} | \text{e}^{-\text{i} \varphi} | \Psi_{\tilde j+1} } |^2 
\;, \label{eqn:unwrapcondition}
\end{equation}
is maximal. Finding this assignment is a $NP$-hard maximization problem. Fortunately, it can be mapped to the traveling salesman problem for which many excellent heuristics were developed \cite{applegate2006traveling}. For finding a solution, we use the Google Optimization Tools \cite{Google_2018}. Conveniently, the maximization of (\ref{eqn:unwrapcondition}) also ensures that eigenstates which belong to consecutive momenta in the reduced Brillouin zone are belonging to consecutive momenta in the extended Brillouin zone as well. Thus, the resulting band structure analogue can be interpreted as the optimally ``unwrapped'' version of the band structure over the reduced Brillouin zone. 
\begin{figure}
\includegraphics{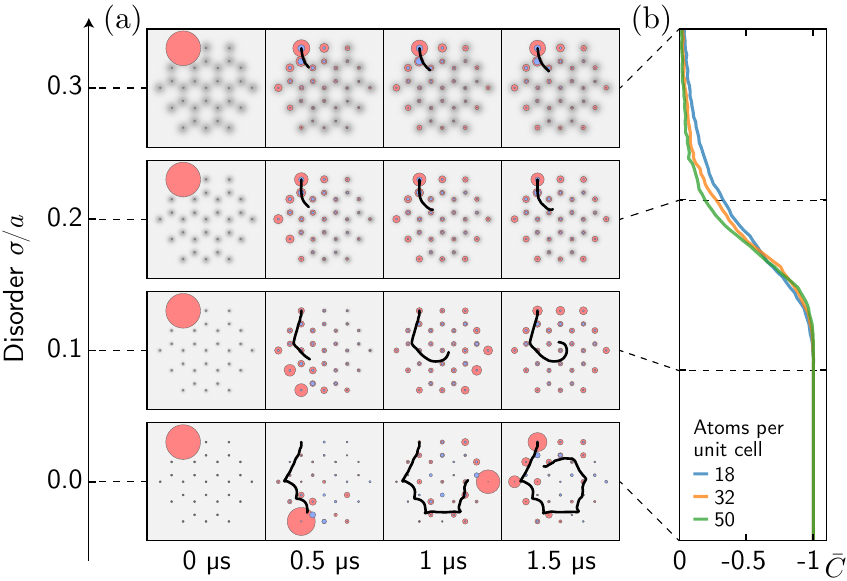}%
\caption{\label{fig:robustness} Effects of lattice disorder. The shifts of the lattice sites obey a normal distribution with standard deviation $\sigma$.  (a) We study the effect on the propagation of an edge excitation averaged over $800$ system realizations. (b) Average Chern number $\bar{C}$ of the lowest band of the infinite honeycomb lattice calculated for unit cells of different sizes for increasing lattice disorder. The standard error is of the size of the width of the plotted curves. The disappearance of the chiral propagation of the edge excitation coincides with the Chern number becoming zero.}
\end{figure}
From the band structure analogue, we see that we have one chiral edge mode even in small system with as few as 10 atoms, see Fig. \ref{fig:edgestates} (b-c). We now have a single edge mode per system because the disk-shaped systems just have one boundary. The dispersion relations of the edge mode are still nearly linear. Thus, we can conclude that the main properties of the edge modes stay the same if we go from semi-infinite systems to small disk shaped systems.

In order to probe the edge modes experimentally, we propose to excite one single atom at the edge into the $\ket{+}$ state. As we have tuned our system such that the two lower bands mainly consist of the $\ket{+}$ state, the excitation has a large overlap with the edge mode between the two lower bands. The simulated time evolution of one edge excitation is shown in Fig. \ref{fig:propagation}.  While most of the excitation stays at the edge, the small overlap of the excitation with bulk states leads to a minor spread within the bulk of the system. The center of mass of the excitation performs a chiral movement with a velocity that matches the group velocity extracted from the dispersion relations shown in Fig. \ref{fig:edgestates} \footnote{The excitation propagates anticlockwise in the disk-shaped systems for a positive group velocity because we consider anticlockwise angles to be positive.}. As the dispersion relations are not completely linear, the excitation broadens in time as it can be seen in the simulated time evolution. These analyses even hold for the small disk shaped system with as few as 10 atoms.

\textit{Robustness.} Experimentally, due to the finite temperature and imperfections in the array of tweezers, the atoms are not perfectly positioned on a regular lattice, but show a small random displacement that varies from shot to shot. For this reason, we study the influence of lattice disorder on the propagation of the excitations. Hereto, we add normal-distributed shifts to the positions of the lattice sites. Fig.~\ref{fig:robustness}~(a) shows the  sample-averaged propagation of edge excitations as a function of the standard deviation $\sigma$ of the positions of the lattice sites. The chiral propagation is robust against the disorder up to $\sigma/a \sim 0.1$. For comparison, we calculated the sample-averaged Chern number $\bar{C}$  of the lowest band of an infinite honeycomb lattice for increasing disorder. To include the disorder into the calculation, we enlarged the unit cell of the honeycomb lattice and added random shifts to the positions of the atoms within the enlarged unit cells. Fig. \ref{fig:robustness} (b) shows $\bar{C}$, which can take non-integer values because of the averaging over several system realizations, for various sizes of the unit cell. For infinite unit cells, we expect a sharp transition between the topological and the trivial phase due to self-averaging of disorder.
The disappearance of the chiral propagation of the edge excitation coincides with the Chern number becoming zero, which confirms nicely that the chiral edge modes are the characteristic feature of the non-zero Chern numbers. A similar analysis indicates that lattice vacancies are tolerable up to a vacancy probability of $\sim 20\%$.
Note that the robustness to vacancies has previously been studied in \cite{Peter_2015, Bettles_2017}.

Due to the robustness, the requirements for realizing such systems are readily met by recent experiments where the lattice disorder is below $\sigma/a < 0.1$ and fully loaded systems with up to $\sim50$ atoms exist \cite{Barredo_2016, Endres_2016, Kim_2016}. Because of the robustness to lattice vacancies, we can also tolerate errors in the preparation of the Rydberg states.

\textit{Experimental realization.} In the following, we give realistic experimental parameters for realizing the Rydberg level structure and hopping amplitudes which were used throughout the paper. We suggest to use the principal quantum number $n=63$ for the $\ket{0}$ state and $n'=62$ for the $\ket{\pm}$ states. These three states can be well isolated from the other Rydberg states and $\mu$ can be tuned to our needs by applying an electric field of $\unit[600]{mV/cm}$ and a magnetic field of $\unit[-15.8]{G}$ along the quantization axis. For a lattice constant of $a=\unit[10]{\mu m}$, we obtain the applied hopping amplitudes as shown in Appendix \ref{sec:tunnelings}.
Note that our setup would work with somewhat different lattice spacing and quantum numbers just as well if the fields are adapted accordingly.

\textit{Conclusion and outlook.} We proposed a Rydberg system which gives rise to topologically protected edge states through dipolar interaction and identified realistic experimental parameters for the implementation of the system. Our proposal has the advantage of low experimental requirements. Signatures of topologically protected edge states exist already in tiny systems with as few as 10 atoms and their demonstrated robustness guaranties that lattice disorder or errors in the preparation of Rydberg states can be tolerated. Thus, our proposal is perfectly suited for recent Rydberg setups which acquired the capability of addressing single Rydberg atoms~\cite{de_L_s_leuc_2017}. We expect our setup to be promising as a starting point for the implementation of bosonic fractional Chern insulators as contrary to many other setups, we do not need time-dependent fields for realizing topological bands. This eliminates a possible source of energy entry into the system which is considered to prevent the experimental realization of fractional Chern insulators.

\begin{acknowledgments}
This research has received funding from the European Research Council (ERC) under the European Union’s Horizon 2020 research and innovation programme (grant agreement No 681208). We acknowledge support from the R\'egion \^Ile-de-France, from the Labex PALM (Xylos project), and from the ``Fondation d'entreprise iXcore pour la Recherche''.
\end{acknowledgments}

\appendix
\section{Microscopic derivation of the hopping amplitudes}\label{sec:tunnelings}

In the following, we review the derivation of the interaction operator (\ref{eqn:interaction}) and calculate the hopping amplitudes. 
Given that the quantization axis is the $z$-axis, the dipole-dipole interaction operator for two atoms $i$ and $j$ in the $xy$-plane reads 
\begin{align}
V_{ij}^\text{dd}  = \frac{1}{4 \pi \epsilon_0 |\boldsymbol{R}_{ij}|^3} \biggl[d_i^{0} d_j^{0}  + \frac{1}{2} \left( d_i^{+} d_j^{-} + d_i^{-} d_j^{+} \right)  \nonumber\\ 
-\frac{3}{2} \left(d_i^{-} d_j^{-} \text{e}^{2\text{i}\phi_{ij}} + d_i^{+} d_j^{+} \text{e}^{-2\text{i}\phi_{ij}} \right) \biggr]
\;,
\label{eqn:dipoledipole}
\end{align}
where $\boldsymbol{R}_{ij} = \boldsymbol{R}_j - \boldsymbol{R}_i$ is the distance vector between the atoms and $\phi_{ij}$ is the angle between the distance vector and the $x$-axis. The operators $d^0 = e z$ and  $d^\pm = e r\sqrt{\frac{4\pi}{3}}\, Y_{1,\pm1}(\vartheta, \varphi) $ are the electric dipole operators, where $e$ is the elementary charge and $Y_{1,\pm1}$ are spherical harmonics. By neglecting static dipole-dipole interactions $d_i^{0} d_j^{0}$ and introducing the hopping amplitudes
\begin{align}
t_+ &= - \frac{1}{\kappa} \braket{ +_i ~ 0_j | d_i^{+} d_j^{-} | 0_i ~ +_j }
= \frac{1}{\kappa}  |\braket{ + | d^{+} | 0  }|^2\;, \nonumber\\
t_- &= - \frac{1}{\kappa}  \braket{ -_i ~ 0_j | d_i^{-} d_j^{+} | 0_i ~ -_j }
= \frac{1}{\kappa}   |\braket{ - | d^{-} | 0  }|^2\;, \nonumber\\
w &= - \frac{3}{\kappa}  \braket{ +_i ~ 0_j | d_i^{+} d_j^{+} | 0_i ~ -_j } = \frac{3}{\kappa} \braket{ + | d^{-} | 0  } \braket{ - | d^{+} | 0  }\;,
\end{align}
with $\kappa=8 \pi \epsilon_0 a^3$, we can transform the depicted dipole-dipole interaction operator~(\ref{eqn:dipoledipole}) into the operator~(\ref{eqn:interaction}).

For calculating the hopping amplitudes, we take into account that the applied fields cause a $\sim\unit[7]{\%}$ admixture of other Rydberg states into the states of the V-level structure. Following \cite{Sobelman_1992, Weber_2017}, we evaluate all relevant dipole matrix elements and obtain the hopping amplitudes $t_+=\unit[2.25]{MHz}$, $t_-=\unit[0.84]{MHz}$, and $w=\unit[4.10]{MHz}$, which we used throughout the paper.

\begin{figure}[b]
\includegraphics{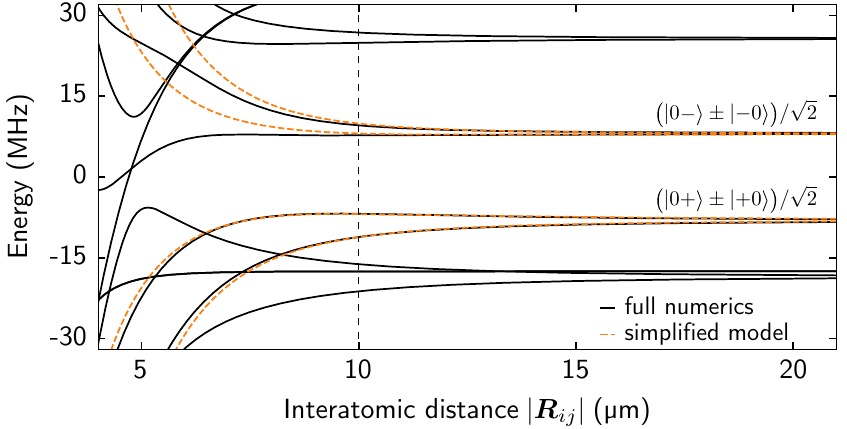}%
\caption{\label{fig:potentials} Pair interaction potentials for $\ket{0}= \ket{63S_{1/2}, m_j=1/2}$, $\ket{+}=\ket{62P_{3/2}, m_j=3/2}$,  $\ket{-}=\ket{62P_{3/2}, m_j=-1/2}$, $E=\unit[600]{mV/cm}$, and $B=\unit[-15.8]{G}$ perpendicular to the interatomic axis. The vertical line marks the lattice constant $a=\unit[10]{\mu m}$. For all relevant distances, the potentials of our simplified model (dashed lines) agree well with the numerically calculated precise potentials (solid lines).}
\end{figure}

\section{Comparison of the interaction potentials of our model with precise potentials}\label{sec:comparison}

As discussed in the main text, our model treats the interaction between a pair of Rydberg atoms in a simplified form. We just consider couplings within the Hilbert space $\{\ket{0}, \ket{+}, \ket{-} \}$, neglecting van der Waals as well as static dipole-dipole interactions. In the following, we show that these simplifications are valid. To this, we compare the pair interaction potentials of the model's Hamiltonian (\ref{eqn:hamiltonian}) with numerically calculated precise interaction potentials. The calculations are carried out as described in \cite{Weber_2017} using recent software \cite{pairinteraction}. The applied basis set comprises $\sim 2000$ Rydberg pair states with $n=59-66$ and $l=0-4$, within an energy range of $\unit[6]{GHz}$. Fig. \ref{fig:potentials} shows the resulting potential curves. The good agreement with the interaction potentials of the model's Hamiltonian justifies the used simplifications.

\end{document}